\documentclass{article}

\usepackage{PRIMEarxiv}

\usepackage[utf8]{inputenc} 
\usepackage[T1]{fontenc}    
\usepackage{hyperref}       
\usepackage{url}            
\usepackage{booktabs}       
\usepackage{amsfonts}       
\usepackage{nicefrac}       
\usepackage{microtype}      
\usepackage{lipsum}
\usepackage{fancyhdr}       
\usepackage{graphicx}       
\graphicspath{{media/}}     
\usepackage{graphicx}
\usepackage{tabularx}
\usepackage{booktabs}
\usepackage{amsmath}
\title{Technical report of a DMD-based Characterization Method for Vision Sensors
}

\author{
  Yapeng Meng, Taoyi Wang, Yihan Lin \\
  Center for Brain-Inspired Computing Research (CBICR) \\
  Tsinghua University \\
  Beijing\\
  \texttt{\{myp23, wangty23, linyh20\}@mails.tsinghua.edu.cn} \\
}

\begin{document}
\maketitle




CMOS image sensors (CIS) convert light intensity on the focal plane pixel-by-pixel and serve as essential sensing devices for various applications, ranging from mobile devices to autonomous driving. The development of pixel architecture, readout circuits, and manufacturing technologies has advanced significantly in recent years, propelling the maturation of characterization standards such as EMVA1288~\cite{jahne2024universal}. This standard utilizes a highly uniform and variable-intensity light source as the inputs stimulus and analyze sensor output under linearly changing illumination conditions. The characterization results including several perspectives.
\begin{itemize}
\vspace{-0.5em}
\item \textbf{Sensitivity}: Quantum efficiency (QE) reflects the proportion of input photons that are converted into charge within a pixel.
\vspace{-0.5em}
\item \textbf{Linearity}: The degree to which the response curve for converting light into digitized values is linear.
\vspace{-0.5em}
\item \textbf{Uniformity}: The dark signal nonuniformity (DSNU) describes the spatial variations without input illumination, while the photo response nonuniformity (PRNU) is the spatial variation with a suitable input illumination.
\item \textbf{Dynamic range}: Dynamic range (DR) in vision sensors is defined as the ratio of the maximum (saturation) to minimum (noise level) detectable illumination levels where signal-to-noise ratio (SNR) > 0dB.

\end{itemize}

The establishment of such a standardized testing system has significantly promoted the standardization of image sensor applications, allowing customers to comprehensively and fairly evaluate the performance of different sensors.

Traditional image sensors densely sample light information in both spatial and temporal domain, making the entire imaging and processing systems are power- and bandwidth-hungry. Thus, brain-inspired vision sensors (BVS), such as silicon retinas and event-based vision sensors (EVS), have been proposed to address the aforementioned limitations by mimicking the human retina. Among the BVS, EVS\cite{Prophesee_IMX636ES_2021, kodama20231, guo20233, iniVation_DAVIS346_2019, iniVation_WhitePaper_2020} are commercially available and have gained significant interest due to their sparse output, low latency, and high dynamic range. The sampling strategies of EVS are asynchronously output ±1-bit events based on temporal changes of contrast. Several companies, including Sony, Prophesee, Samsung, and OmniVision, are actively developing event camera chips. Additionally, recent work has proposed hybrid chips that combine both RGB and EVS technologies~\cite{kodama20231, guo20233, iniVation_DAVIS346_2019}. However, due to their inherent characteristics, such as responding only to dynamic scenes and losing features when the camera moving parallel to the edges, EVS cannot be characterized by well-established standards such as EMVA1288. Currently, the mainstream testing methods used in EVS, including objective observation for DR test~\cite{lichtsteiner2008128}, integrating sphere tests with varying light sources~\cite{mcreynolds2022experimental, moeys2017sensitive}, and direct test the logarithmic pixel response without event circuits~\cite{schon2023320}. As a result, even if the same chip is used, different test results will be obtained under different test methods. 

The fundamental difference between BVS and CIS lies in their output characteristics, making it challenging to fully adapt the EMVA1288 standard for their characterization. BVS outputs are directly influenced by input variations, such as the rate of light change or the richness of edge textures. This characteristic makes it impractical to use spatially or temporally consistent light sources for testing. Therefore, an advanced optical testing system that can dynamically control light intensity outputs in both spatial and temporal domains, at high speed and with a wide dynamic range, is essential.

Recently, an emerging class of BVS has been proposed. Inspired by the human visual system (HVS), which decomposes complex environments into visual primitives and processes them via complementary data pathways, Tianmouc has been developed to address the visual challenge in open-world environments ~\cite{yang2024vision}. This chip integrates two complementary pathways: a cognition-oriented pathway (COP) that captures color intensity, and an action-oriented pathway (AOP) that provides multi-bit spatial differences (SD) and multi-bit temporal differences (TD), thereby obtaining comprehensive visual information. Benefit from the global operation in capturing TD, SD, and color intensity, the characterization of the Tianmouc chip can adhere to well-established standards. For instance, the high-precision quantization of SD and TD enables the definition, computation, and measurement of its signal-to-noise ratio (SNR) in a manner similar to the EMVA1288 standard. However, the testing of Tianmouc remains highly cumbersome. Direct testing of the SD requires precise pixel-level spatial alignment, while testing the TD demands accurate light source control. Like the EVS, Tianmouc also anticipates having a standardized and user-friendly testing platform.

\begin{figure*}
    \centering{\includegraphics[width=0.8\linewidth]{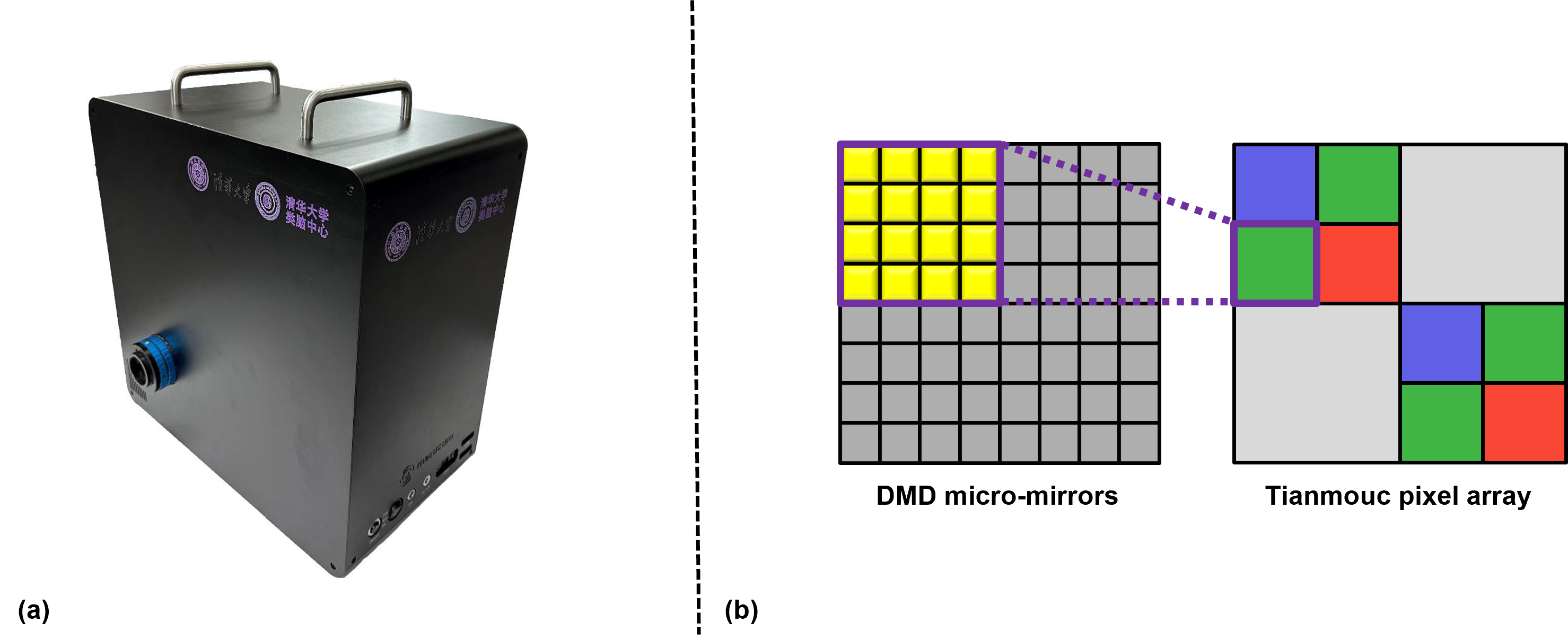}}\\
     \caption{(a) The physical drawing of our equipment. (b) A schematic of multi-pixel spatial modulation. We use multiple micro-mirrors corresponding to one Tianmouc pixel.}
    \label{fig:1}
\end{figure*}

\section*{A Dmd-based Characterization Method}
A digital micromirror device (DMD)-based testing approach presents a promising solution. These devices are widely used in applications such as projectors and optical computing. A DMD consists of microscopically small mirrors arranged in a matrix on a semiconductor chip, where each mirror represents a pixel in the projected image and can rapidly switch between “on” and “off” states to modulate light reflection. In the temporal domain, higher bit-depth can be achieved by accumulating data over a longer period. Similarly, in the spatial domain, faster encoding can be realized through multi-pixel spatial modulation, as illustrated in the figure~\ref{fig:1} (b). With its high temporal resolution, a DMD enables the testing of asynchronous ±1-bit temporal contrast changes in EVS and ±7-bit TD in Tianmouc. Additionally, its high spatial resolution allows for precise testing of ±7-bit SD in Tianmouc. Furthermore, it is well known that EVS latency is directly related to the number of output events, spatial distribution of events, and the rate of input changes. Traditional EVS testing methods rely on varying light sources can only evaluate a limited number of EVS pixels. In contrast, the DMD-based testing approach enables reproducible and quantitative analysis of the relationship between the number of EVS events, different input stimuli, and EVS output. This method also allows for a quantitative assessment of event rate saturation and other limitations of current EVS technology. Moreover, since EVS response speed is also influenced by light intensity, the DMD-based approach provides a more precise way to analyze the relationship of intensity and latency.
\begin{figure*}
    \centering{\includegraphics[width=0.8\linewidth]{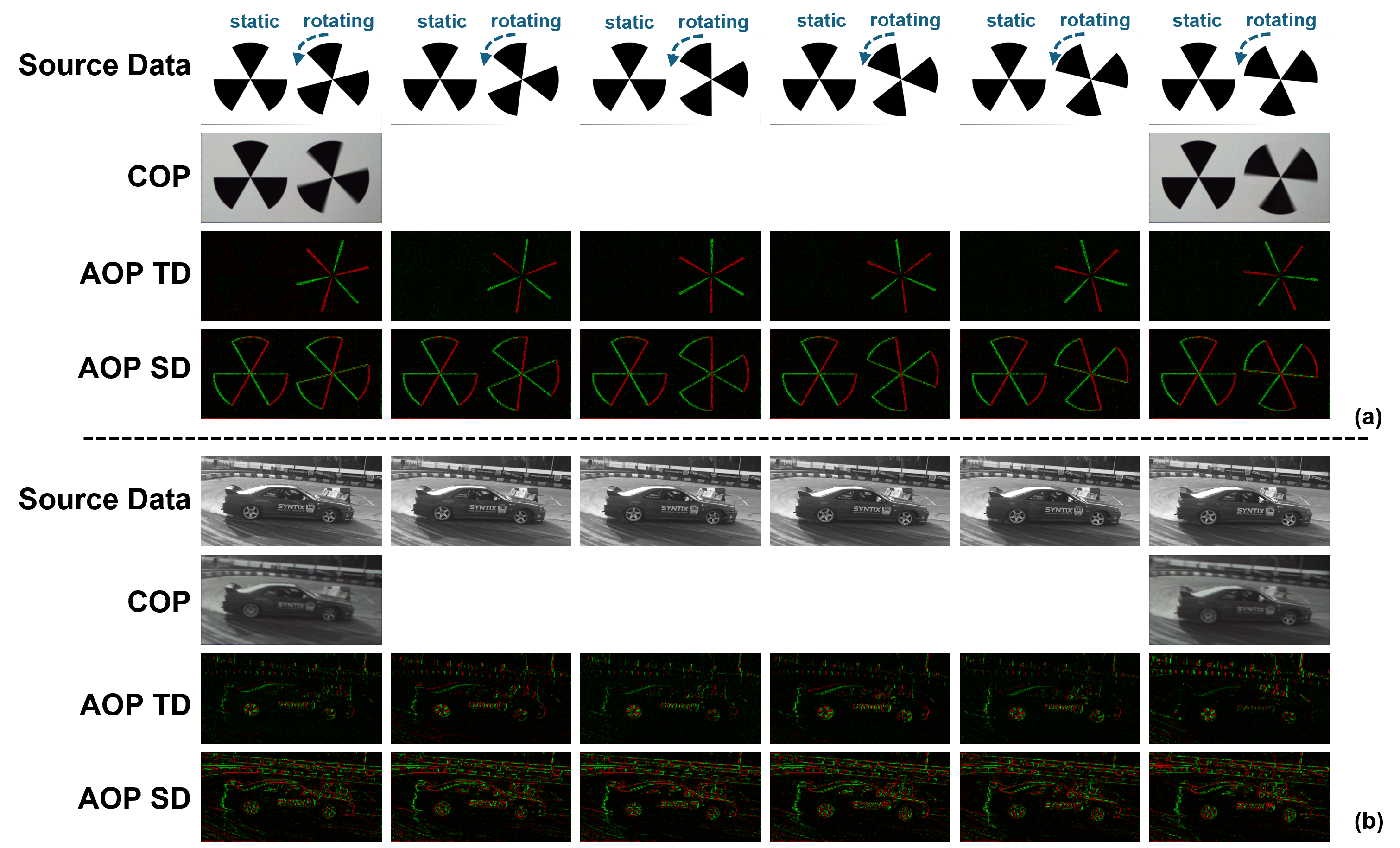}}\\
     \caption{Visualization results of our characterization method. (a) and (b) respectively show the projection results of a specific simple image pattern and a real-world scene onto the Tianmouc camera using a DMD chip. Due to the high frame rate of the AOP pathway in the Tianmouc chip, in the 757 FPS mode, there are 25 AOP frames between two consecutive COP frames. In the figure, one AOP frame is selected for display every 5 frames. }
    \label{fig:result}
\end{figure*}

Since Tianmouc currently provides the most comprehensive information of neuromorphic sensors, including color intensity, ±7-bit SD, and ±7-bit TD, we have developed an optical testing system using Tianmouc as a representative reference. We utilize the Texas Instruments TI5090 DMD. To ensure high-speed, high-accuracy, and high dynamic range optical projection, we employ a spatiotemporal encoding strategy. This strategy maps multiple 4×4 DMD mirrors to a single COP pixel of Tianmouc with a 4 $\mu m$ pixel pitch and 8×8 DMD mirrors to a single AOP pixel of Tianmouc with an 8 $\mu m$ pixel pitch. By synchronizing the DMD micromirror flipping with the sensor's exposure through hardware-level triggers, we achieve temporal alignment. Furthermore, through optical path design, we ensure the alignment of the reflected light from the DMD with the sensor's pixels, achieving spatial alignment. Based on the above design, a chip characterization method with high precision and high speed can be achieved through temporal-spatial alignment.

Figure~\ref{fig:result} presents the visualization results of our characterization method. In Figure~\ref{fig:result}(a), a specific rotating image pattern is shown. The left-side pattern remains static, while the right-side pattern rotates around its center. In the projection results, the left-side pattern appears black in the AOP TD channel, whereas the right-side pattern exhibits motion edges. In the AOP SD channel, the spatial edge information of both patterns is clearly presented. Figure~\ref{fig:result}(b) shows a real-world dynamic scene where both the camera and the car are in motion. It can be observed that for regions with intense motion, such as the car wheels, the AOP TD channel exhibits a stronger response.

\section*{Discussion and Future Prospects}
The main challenge for this characterization method is the complexity of the optical system. The lenses must be customized to accommodate differences in pixel size between sensors, resulting in higher costs. 

This characterization method also holds promise for large-scale generation and conversion of BVS datasets. Similar to the effect shown in Figure~\ref{fig:result}(b), a color image can be split into R, G, and B channels, which are then projected onto the BVS chip separately using this approach. The per-channel projection results are subsequently combined into the COP and AOP pathways, enabling the transformation of various image or video datasets into corresponding BVS datasets.


\bibliographystyle{unsrt}  
\bibliography{references}

\begin{thebibliography}{10}

\bibitem{jahne2024universal}
Bernd J{\"a}hne.
\newblock A universal concept to characterize modern image sensors.
\newblock In {\em Forum Bildverarbeitung 2024}, page~13. KIT Scientific Publishing, 2024.

\bibitem{Prophesee_IMX636ES_2021}
Prophesee.
\newblock Imx636es (hd), 2021.

\bibitem{kodama20231}
Kazutoshi Kodama, Yusuke Sato, Yuhi Yorikado, Raphael Berner, Kyoji Mizoguchi, Takahiro Miyazaki, Masahiro Tsukamoto, Yoshihisa Matoba, Hirotaka Shinozaki, Atsumi Niwa, et~al.
\newblock 1.22 $\mu$ m 35.6 mpixel rgb hybrid event-based vision sensor with 4.88 $\mu$m-pitch event pixels and up to 10k event frame rate by adaptive control on event sparsity.
\newblock In {\em 2023 IEEE International Solid-State Circuits Conference (ISSCC)}, pages 92--94. IEEE, 2023.

\bibitem{guo20233}
Menghan Guo, Shoushun Chen, Zhe Gao, Wenlei Yang, Peter Bartkovjak, Qing Qin, Xiaoqin Hu, Dahei Zhou, Masayuki Uchiyama, Yoshiharu Kudo, et~al.
\newblock A 3-wafer-stacked hybrid 15mpixel cis+ 1 mpixel evs with 4.6 gevent/s readout, in-pixel tdc and on-chip isp and esp function.
\newblock In {\em 2023 IEEE International Solid-State Circuits Conference (ISSCC)}, pages 90--92. IEEE, 2023.

\bibitem{iniVation_DAVIS346_2019}
iniVation.
\newblock {\em DAVIS 346}, 2019.

\bibitem{iniVation_WhitePaper_2020}
iniVation.
\newblock Understanding the performance of neuromorphic event-based vision sensors.
\newblock Technical report, iniVation, 2020.

\bibitem{lichtsteiner2008128}
Patrick Lichtsteiner, Christoph Posch, and Tobi Delbruck.
\newblock A 128 $times $128 120 db 15 $mu $ s latency asynchronous temporal contrast vision sensor.
\newblock {\em IEEE journal of solid-state circuits}, 43(2):566--576, 2008.

\bibitem{mcreynolds2022experimental}
Brian McReynolds, Rui Graca, and Tobi Delbruck.
\newblock Experimental methods to predict dynamic vision sensor event camera performance.
\newblock {\em Optical Engineering}, 61(7):074103--074103, 2022.

\bibitem{moeys2017sensitive}
Diederik~Paul Moeys, Federico Corradi, Chenghan Li, Simeon~A Bamford, Luca Longinotti, Fabian~F Voigt, Stewart Berry, Gemma Taverni, Fritjof Helmchen, and Tobi Delbruck.
\newblock A sensitive dynamic and active pixel vision sensor for color or neural imaging applications.
\newblock {\em IEEE transactions on biomedical circuits and systems}, 12(1):123--136, 2017.

\bibitem{schon2023320}
Guillaume Schon, Denis Bourke, Pierre-Antoine Doisneau, Thomas Finateu, Adrien Gonzalez, Naoyuki Hanajima, Tahar Hitana, Lucas~Janse Van~Vuuren, Moataz Kadry, Charles Laurent, et~al.
\newblock A 320 x 320 1/5” bsi-cmos stacked event sensor for low-power vision applications.
\newblock In {\em 2023 IEEE Symposium on VLSI Technology and Circuits (VLSI Technology and Circuits)}, pages 1--2. IEEE, 2023.

\bibitem{yang2024vision}
Zheyu Yang, Taoyi Wang, Yihan Lin, Yuguo Chen, Hui Zeng, Jing Pei, Jiazheng Wang, Xue Liu, Yichun Zhou, Jianqiang Zhang, et~al.
\newblock A vision chip with complementary pathways for open-world sensing.
\newblock {\em Nature}, 629(8014):1027--1033, 2024.

\end{thebibliography}

\end{document}